%% file: sphoton.tex
\documentclass[english,superscriptaddress,showpacs,showkeys]{revtex4-2}
\usepackage[T1]{fontenc}
\usepackage[utf8]{inputenc}
\usepackage{color}
\usepackage{babel}
\usepackage{array}
\usepackage{units}
\usepackage{amsmath}
\usepackage{amssymb}
\usepackage{graphicx}
\usepackage{bm}
\usepackage{subcaption}
\usepackage{todonotes}
\usepackage[pdfusetitle,
bookmarks=true,bookmarksnumbered=false,bookmarksopen=true,bookmarksopenlevel=2,
breaklinks=true,pdfborder={0 0 0},pdfborderstyle={},backref=false,colorlinks=true]
{hyperref}
\hypersetup{
	citecolor=blue,linkcolor=blue,urlcolor=blue}


\input{commands}

\begin{document}
	
	\title{Soft photon approximation in a laser field}
	
	\author{P.A. Krachkov}\email{P.A.Krachkov@inp.nsk.su}
	\affiliation{Budker Institute of Nuclear Physics, SB RAS, Novosibirsk, 630090, Russia}
	\affiliation{Novosibirsk State University, 630090 Novosibirsk, Russia}
	
	\date{\today}
\begin{abstract}
In the present paper, we consider processes involving the emission of soft photons in the presence of a strong laser field. We  demonstrate that the matrix element $S$ for a process $\text{i} \rightarrow \text{f} + \gamma$, with a soft photon $\gamma$, can be expressed in terms of the matrix element $S_0$ for the process $\text{i} \rightarrow \text{f}$ through a simple multiplicative factor \eqref{eq:FGpm} in the integrand over $\phi$. This approximation enables a result that is exact in the phase and approximate in the prefactor to order $\mathcal{O}(\omega/\varepsilon_\text{char})$, where $\omega$ is the frequency of the soft photon and $\varepsilon_\text{char}$ is the characteristic energy of the $\text{i}\rightarrow\text{f}$ process.

We demonstrate several important applications of this soft photon approximation. First, under soft photon approximation  we compute the probabilities of nonlinear Compton scattering and photon emission in the superposition of a laser and atomic fields and compare obtained result with the exact one. Second, we demonstrate that the amplitude of $n$ soft photons emission has factorization, which corresponds to the independence of the emission of $n$ soft photons. Third, we use the discussed approximation to prove cancellation of real and virtual infrared divergences for nonlinear Compton scattering and derive the finite radiative corrections.

The soft photon approximation is a useful tool for investigation of different QED processes  in the presence of a strong laser field. Also, it can be widely used for computation of infrared part of radiative correction for some processes.

\end{abstract}
	
\maketitle

\section{Introduction}\label{sec:intr}
High-intensity laser pulses, enabled by chirped pulse amplification~\cite{Strickland1985}, have great potential as a tool to probe quantum electrodynamics (QED) in intense background fields. The presence of intense background electromagnetic fields open up new opportunities for experimental and theoretical studies of QED in the nonlinear strong-field regime, where background fields strongly affect the physical processes and dynamics of charged particles. 

The theoretical description of basic strong-field QED processes like nonlinear Compton scattering and nonlinear Breit-Wheeler pair production has been studied  in detail by approximating the laser field as a plane wave; see reviews \cite{Mitter1975,Ritus1985,Baier1998,Ehlotzky2009, DiPiazza2012, Fedotov2022a}. The results of  computations are much more difficult structure than the corresponding result in the vacuum case, even for these $1\rightarrow2$ processes (nonlinear Compton scattering and Breit-Wheeler pair production). The processes  $1\rightarrow3$ and $2\rightarrow2$, such as double nonlinear Compton scattering, electron trident production, electron-positron annihilation into two photons, etc., have a much more complex structure of theoretical results compared to the $1\rightarrow2$ processes in the presence of intense laser fields. 

The complexity of the results increases exponentially with the number of particles in the initial and final states. Thus, developing approximate methods is critical. They allow for deriving new results and analyzing and verifying the exact results. There currently exist several approximate methods for QED processes in intense laser fields, such as the local constant field approximation \cite{Nikishov1985}, the quasiclassical approximation \cite{BK1968}, see also reviews \cite{DiPiazza2012,Fedotov2022a}. It is well known that the lower the photon emission energy, the higher the emission probability. Therefore, studying processes involving soft photon emission is essential.

The present work is devoted to studying this problem. We demonstrate that the matrix element $S$ for the process $\text{i}\rightarrow\text{f}+\gamma$ involving a soft photon $\gamma$ can be expressed in terms of the  matrix element $S_0$ of the process $\text{i}\rightarrow\text{f}$, where $\text{i}$ and $\text{f}$ denote the set of initial and final particles. This approximation is an analogue of the soft photon approximation for studying QED processes in the absence of strong external fields. This substantially simplifies the matrix element, making it an approximation of the prefactor while retaining an exact phase. We discuss several simple applications of the soft photon approximation. We also examine the factorization of $n$-photon emission amplitudes and the cancellation between real and virtual infrared divergences.

\section{Amplitude of soft photon emission}\label{sec::ampl}

Let $d W_0$ be the probability (cross section) for a given "hard" process $\text{i}\rightarrow\text{f}$ of charged particles in the presence of a strong laser field, which may be accompanied by the emission of a certain number of photons. Where  $\text{i}$ and $\text{f}$ denote the set of initial and final particles. Together with this process, we will consider another that differs from it only in that one extra photon is emitted. If the frequency $\omega$ of this photon is sufficiently small (the necessary conditions will be formulated below), the probability $dW$ for the second process ($\text{i}\rightarrow\text{f}+\gamma$) is related in a simple manner to $dW_0$. In this case, we can neglect the influence of the emission of this quantum on the $\text{i}\rightarrow\text{f}$ process. The probability $dW$ can therefore be simply represented through $dW_0$ and the  probability $dI$ of single photon  emission  in the collision in the presence of a strong laser field. 

The diagrams for the process involving an additional photon are obtained from those for the original process by adding an external photon line that "branches off" for an (external or internal) electron line. It is easily seen that the most important diagrams will be those in which this change is made in external electron lines. 
If $p$ and $k$ are the momenta of an external electron line  and soft photon, the Green's function $G(p\pm k)$ added to the diagram is near the pole for small $\omega=k^0$. That is, when a photon is emitted from an initial or final electron, it has a large formation length \cite{BAIER2005261,Ritus1985}. However, for photon emission from an internal electron line, the formation length is restricted by the hard sub-processes in the diagram \footnote{Resonances exist in many QED processes in the presence of a long laser pulse. These resonances relate to the electron in the intermediate state being near the mass surface, which leads to a cascade. The radiation in the intermediate state will not be suppressed in this cascade. However, the emission of soft photons in these cascade processes falls outside the scope of this article and will not be considered.}.  

The matrix element of the "hard" processes $S_0$ can be represented in the following manner
\begin{equation}\label{eq:M0}
S_0=\int d^4 x\bar{U}^{(out)}_{p'}(x)\,\hat{O}(P,X)\,U^{(in)}_{p}(x)
\,,
\end{equation}
where $\hat{O}$ is an operator, which depends on the type of a process, $P$ and $X$ are operators, and $U_{p}(x)$ is an  electron wave function in the presence of a laser field (Volkov's solution) with an asymptotically four-momentum $p$, see Eq.~\eqref{eq:vawe_func}.  See also \ref{sec:app} for definition and  useful formulae.    Here, for simplicity, we assume that in the  process there is one initial and one final electron. The results will be summarized below for the case of several charged particles in the initial and final states. Matrix element $S$ can be represented as a sum $S=S_1+S_2$, where $S_1$ and $S_2$ correspond to photon emission from initial and final states, respectively. We neglect the contribution of diagrams related to radiation in intermediate states.   Matrix elements $S_{1,2}$ have the following form
\begin{align}\label{eq:M12}
S_1&=e\int d^4 x\bar{U}^{(out)}_{p'}(x)\,\hat{O}(P,X)\frac{1}{[\hat{\Pi}(\Phi)]^{2}-m^{2}+i0}[\hat{\Pi}(\Phi)+m]e^{ikX}\hat{e}^{*}U^{(in)}_{p}(x)\,,\nonumber\\
S_2&=e\int d^4 x\bar{U}^{(out)}_{p'}(x)\,e^{ikX}\hat{e}^{*}[\hat{\Pi}(\Phi)+m]\frac{1}{[\hat{\Pi}(\Phi)]^{2}-m^{2}+i0}\hat{O}(P,X)U^{(in)}_{p,}(x)\,,
\end{align}
where operator $\hat{O}$ is the same as in Eq.~\eqref{eq:M0}, $e$ is the electron charge,  $e^\mu$ and $k^\mu$ are photon polarization and momentum vector, and $\Pi^{\mu}(\Phi)=P^{\mu}-eA^{\mu}(\Phi)$ with $P^{\mu}=i\partial^{\mu}$.

In order to simplify the expression for $S_{1}$ under the soft photon assumption, let us consider the quantity 
\begin{align}
\scalebox{1.64}{$\tau$}  &=\frac{1}{[\hat{\Pi}(\Phi)]^{2}-m^{2}+i0}[\hat{\Pi}(\Phi)+m]e^{ikX}\hat{e}^{*}U^{(in)}_{p}(x)=e^{ikX}\frac{1}{[\hat{\Pi}(\Phi)-\hat{k}]^{2}-m^{2}+i0}[\hat{\Pi}(\Phi)-\hat{k}+m]\hat{e}^{*}U^{(in)}_{p}(x)\nonumber\\
&=e^{ikX}\frac{1}{[\hat{\Pi}(\Phi)-\hat{k}]^{2}-m^{2}+i0}\left[2\pi_{p,\lambda}(\phi)+i\frac{e\hat{n}\hat{A}'(\phi)}{p_{-}}n_{\lambda}-\hat{k}\gamma_{\lambda}\right]e^{*\lambda}U^{(in)}_{p}(x)\,.
\end{align}
Here we commute operators $\hat{\Pi}$ and $\hat{e}^{*}$ and  use Dirac equation and the following identities
\begin{equation}
\Pi^{\lambda}(\phi)U^{(in)}_{p}(x)=\left[\pi_{p}^{\lambda}(\phi)+i\frac{e\hat{n}\hat{A}'(\phi)}{2p_{-}}n^{\lambda}\right]U^{(in)}_{p}(x),
\end{equation}
where 
\begin{equation}
\pi_{p}^{\lambda}(\phi)=p^{\lambda}-eA^{\lambda}(\phi)+\frac{e(pA(\phi))}{p_{-}}n^{\lambda}-\frac{e^{2}A^{2}(\phi)}{2p_{-}}n^{\lambda}\,,
\end{equation}
is the classical kinetic four-momentum of an electron in the plane wave $A^{\mu}(\phi)$, with $\lim_{\phi\rightarrow\pm\infty}\pi_{p}^{\lambda}(\phi)=p^{\lambda}$.
Note that $\pi_{p}(\phi)^{2}=p^{2}$ and $\pi_{p-}(\phi)=p_{-}$.

Using the integral representation of the squared electron propagator Eq.~\eqref{eq:G} and identities Eqs.~\eqref{eq:id1}, \eqref{eq:id2}, we obtain the following result
\begin{equation}
\begin{split} \scalebox{1.64}{$\tau$} & =-ie^{ikx-i(p_{-}T-\bm{x}_{\perp}\cdot\bm{p}_{\perp})}\int_{0}^{\infty}due^{-im^{2}u}\left\{ 1-\frac{e\hat{n}[\hat{A}(\phi_{u})-\hat{A}(\phi)]}{2(p_{-}-k_{-})}\right\} e^{-i\int_{0}^{u}du'[\bm{p}_{\perp}-\bm{k}_{\perp}-e\bm{A}_{\perp}(\phi_{u'})]^{2}} \\
&\times  e^{-2iu(p_{-}-k_{-})(P_{\phi}+k_{+})}\left[2\pi_{p,\lambda}(\phi)+i\frac{e\hat{n}\hat{A}'(\phi)}{p_{-}}n_{\lambda}-\hat{k}\gamma_{\lambda}\right]e^{*\lambda}U^{(in)}_{p}(\phi)\\
& =-ie^{ikx-i(p_{-}T-\bm{x}_{\perp}\cdot\bm{p}_{\perp})}\int_{0}^{\infty}due^{-im^{2}u}\left\{ 1-\frac{e\hat{n}[\hat{A}(\phi_{u})-\hat{A}(\phi)]}{2(p_{-}-k_{-})}\right\} e^{-i\int_{0}^{u}du'[\bm{p}_{\perp}-\bm{k}_{\perp}-e\bm{A}_{\perp}(\phi_{u'})]^{2}}\\
&\times e^{-2iu(p_{-}-k_{-})k_{+}}\left[2\pi_{p,\lambda}(\phi_{u})+i\frac{e\hat{n}\hat{A}'(\phi_{u})}{p_{-}}n_{\lambda}-\hat{k}\gamma_{\lambda}\right]e^{*\lambda}U^{(in)}_{p}(\phi_{u}) \,,
\end{split}
\end{equation}
where $\phi_{u}=\phi-2u(p_{-}-k_{-})$ and the function $U^{(in)}_{p}(\phi)=U^{(in)}_{p}(x)e^{i(p_{-}T-\bm{x}_{\perp}\cdot\bm{p}_{\perp})}$ depends only on $\phi$. The quantity $U^{(in)}_{p}(\phi_u)$ can be expressed through  $U^{(in)}_{p}(\phi)$ by the identity
\[
U_{p}^{(in)}(\phi_{u})=\left\{ 1+\frac{e\hat{n}[\hat{A}(\phi_{u})-\hat{A}(\phi)]}{2p_{-}}\right\} e^{i\left\{ -p_{+}(\phi_{u}-\phi)-\int_{\phi}^{\phi_{u}}d\phi'\left[-\frac{e\bm{p}_{\perp}\cdot\bm{A}_{\perp}(\phi')}{p_{-}}+\frac{e^{2}\bm{A}_{\perp}^{2}(\phi')}{2p_{-}}\right]\right\} }U_{p}^{(in)}(\phi)\,.
\]

The final result has the following form
\begin{equation}
\begin{split} \scalebox{1.64}{$\tau$}  &=-ie^{ikX}\int_{0}^{\infty}due^{-im^{2}u}e^{i\int_{0}^{u}du'[\bm{\pi}_{p}(\phi_{u}')-\bm{k}]^{2}}\left\{ 1-\frac{e\hat{n}[\hat{A}(\phi_{u})-\hat{A}(\phi)]}{2(p_{-}-k_{-})}\right\} \\
&\times \left[2\pi_{p,\lambda}(\phi_{u})+i\frac{e\hat{n}\hat{A}'(\phi_{u})}{p_{-}}n_{\lambda}-\hat{k}\gamma_{\lambda}\right]e^{*\lambda}\left\{ 1+\frac{e\hat{n}[\hat{A}(\phi_{u})-\hat{A}(\phi)]}{2p_{-}}\right\} U^{(in)}_{p}(x)=\\
& =-ie^{ikX}\int_{0}^{\infty}due^{-2i\int_{0}^{u}du'(\pi_{p}(\phi_{u}')k)}\left\{ 1-\frac{e\hat{n}[\hat{A}(\phi_{u})-\hat{A}(\phi)]}{2(p_{-}-k_{-})}\right\} \\
&\times \left[2\pi_{p,\lambda}(\phi_{u})+i\frac{e\hat{n}\hat{A}'(\phi_{u})}{p_{-}}n_{\lambda}-\hat{k}\gamma_{\lambda}\right]e^{*\lambda}\left\{ 1+\frac{e\hat{n}[\hat{A}(\phi_{u})-\hat{A}(\phi)]}{2p_{-}}\right\} U^{(in)}_{p}(x)\,.
\end{split}
\label{eq:res1}
\end{equation}
Here we use a method similar to \cite{PhysRevD.102.076018}. This is an accurate result because, up to this point, we have not yet used the smallness of $\omega$. To obtain the result for $k\ll p$, we can neglect a term proportional to $k^{\mu}$ in the preexponential factor. It should be noted that the quantity  $\frac{dA^{\mu}(\phi_{u})}{d\phi_{u}}$ can be expressed in the form
$$\frac{dA^{\mu}(\phi_{u})}{d\phi_{u}}=\frac{-1}{2(p_{-}-k_{-})}\frac{d\left(A^{\mu}(\phi_{u})-A^{\mu}(\phi)\right)}{du}\,,$$
and that following integration by parts, this term will be proportional to $k$ and can also be neglected.  Thus, the final result has the following form
\begin{equation}
\scalebox{1.64}{$\tau$}=-2ie^{ikX}\int_{0}^{\infty}due^{-2i\int_{0}^{u}du'(\pi_{p'}(\phi_{u'})k)}(\pi_{p'}(\phi_{u})e^*)U^{(in)}_{p}(x)\,.
\end{equation}

The amplitudes, which correspond to  the photon emission by the final electron and the initial (final) positron, are obtained in a similar way. The final result for all four cases has the following form
\begin{equation}
\begin{split} & \frac{1}{[\hat{\Pi}(\Phi)]-m+i0}\hat{e}^{*}e^{ikX}U^{(in)}_{p}(x)= F_{p}^{(-)}(\phi)e^{ikX}U^{(in)}_{p}(x)\,,\\
& \bar{U}^{(out)}_{p}(x)e^{ikX}\hat{e}^{*}\frac{1}{[\hat{\Pi}(\Phi)]-m+i0}=F_{p}^{(+)}(\phi)e^{ikX}\bar{U}^{(out)}_{p}(x)\,,\\
& \frac{1}{[\hat{\Pi}(\Phi)]-m+i0}\hat{e}^{*}e^{ikX}V^{(out)}_{p}(x)=G_{p}^{(+)}(\phi)e^{ikX}V^{(out)}_{p}(x)\,,\\
& \bar{V}^{(in)}_{p}(x)e^{ikX}\hat{e}^{*}\frac{1}{[\hat{\Pi}(\Phi)]-m+i0}=G_{p}^{(-)}(\phi)e^{ikX}\bar{V}^{(in)}_{p}(x)\,,
\end{split}
\end{equation}
where
\begin{align}\label{eq:FGpm}
F_{p}^{(-)}(\phi)&=-i\int_{-\infty}^{\phi}\frac{d\varphi}{p_{-}}e^{-i\int_{\varphi}^{\phi}\frac{d\varphi'}{p_{-}-k_-}(\pi_{p}(\varphi')k)}(\pi_{p}(\varphi) e^*)\,,\nonumber\\
F_{p}^{(+)}(\phi)&=-i\int_{\phi}^{+\infty}\frac{d\varphi}{p_{-}}e^{i\int_{\phi}^{\varphi}\frac{d\varphi'}{p_{-}+k_-}(\pi_{p}(\varphi')k)}(\pi_{p}(\varphi) e^*)\,,\nonumber\\
G_{p}^{(-)}(\phi)&=-i\int_{-\infty}^{\phi}\frac{d\varphi}{p_{-}}e^{i\int_{\varphi}^{\phi}\frac{d\varphi'}{p_{-}-k_-}(\pi_{-p}(\varphi')k)}(\pi_{-p}(\varphi) e^*)\,,\nonumber\\
G_{p}^{(+)}(\phi)&=-i\int_{\phi}^{+\infty}\frac{d\varphi}{p_{-}}e^{-i\int_{\phi}^{\varphi}\frac{d\varphi'}{p_{-}+k_-}(\pi_{-p}(\varphi')k)}(\pi_{-p}(\varphi) e^*)\,.
\end{align}
Note that the quantity $\pi_{-p}(\varphi) $ that occurs in the positron factor \eqref{eq:FGpm} can be represented as $-\pi_{p}(\varphi)$ with the substitution $e\rightarrow-e$, where $e$ is an electron charge.

The matrix elements $S_{1,2}$ have the following form
\begin{align}
S_1&=\int d^4 x\bar{U}^{(out)}_{p'}(x)\hat{O}(P,X)U^{(in)}_{p}(x)F_{p}^{(-)}(\phi)\,,\nonumber\\
S_2&=\int d^4 xF_{p}^{(+)}(\phi)\, \bar{U}^{(out)}_{p'}(x)\hat{O}(P,X)U^{(in)}_{p}(x)\,,\nonumber\\
S&=S_1+S_2=\int d\phi \tilde S_0(\phi) \left(F_{p}^{(-)}(\phi)+F_{p'}^{(+)}(\phi)\right)\,,
\end{align}
where  $S_0=\int d\phi \tilde S_0(\phi)$. Here we assume that  $\tilde S_0$ depends on one phase. In the general case, $\tilde S_0$ should depend on several different phases $\phi_i$. For example, for double Compton scattering, the integrand  $S_0$ depends on two phases. In such a case, different factors will depend on different phases. The characteristic phase difference is $\phi_i-\phi_j\ll\varphi_{char}$, where $\varphi_{char}$ is the formation length of soft photon emission. Thus, we can use the same phase for all factors. The same reasoning holds if we consider the probability of some given processes. After partial integration over the phase space of the "hard" process, the phase difference between the integrands $S_0$ and   $S_0^*$ became $\phi-\phi'\ll\varphi_{char}$.

The recipe for using the soft photon approximation  is as follows. To obtain the amplitude $S$ from the amplitude of the "hard" process $S_0$, we should multiply the integrand  $S_0$  with respect to the variable $\phi$ by the sum of the factors. Where each factor corresponds to the photon emission from a given external charged particle. 

Note that the final result is a gauge invariant with the required accuracy since, when replacing $e^\mu$ with $k^\mu$ in the integrand of $F^{\pm}_p$ we get the full derivative.

The conditions for the applicability of soft photon radiation are as follows. First, we assume that $\omega\ll \varepsilon$ and $k_-\ll p_-$. Another condition is that we can neglect the photon momenta in the amplitude $S_0$. This condition depends on the process and cannot be described in a general way.

It is interesting to compare the obtained results with the classical photon emission \cite{LL4}. If we neglect $k_-$ in comparison with $p_-$ in the phase, the result in \eqref{eq:FGpm} will coincide with the classical one \cite{LL4}, obtained by discontinuously changing current density four-vector. Note that in the approximation under consideration, we neglect terms proportional to $k_\mu$ only in the  preexponential factors. We calculate the phase multiplier exactly. The exact account of $k$ in a phase is very important even in the case of $\omega\ll \varepsilon$ due to the integration of a laser pulse length in phase. The difference in phase in the order of unity significantly changes the probability.  Factors \eqref{eq:FGpm} are independent of the spin of a particle and valid also for particles with arbitrary spin.

\subsection{Special Cases}

\begin{itemize}
\item For the absence of the laser field, factors $F_{p}^{(\pm)}$ have the following form
$$
F_{p}^{(\pm)}(\phi)=\pm\frac{pe^*}{pk}\,,
$$
which coincides with the ordinary soft photon approximation factor \cite{LL4}.

\item In the case of $p_{-}\omega_{0}\ll \pi_p k$, where $\omega_0$ is a laser field frequency, the result is significantly simplified
\begin{equation}\label{eq:F_lcf}
F_{p}(\phi)^{(\pm)}=\pm\frac{\pi_p(\phi)e^* }{\pi_p(\phi)k}\,.
\end{equation}
Which corresponds to the local constant field approximation. Note that in the case under consideration, the process $S_0$ should also be considered in the local constant field approximation.

\item For the high-energy particles counter-propagating to the laser field, we have
\[
k\pi_{p}=\frac{\omega}{2\varepsilon}(m^{2}+(\varepsilon\bm{\theta}{}_{pk}-\bm{A})^{2})\,,\quad\pi_{p}e=-(\varepsilon\bm{\theta}_{pk}-e\bm{A})\cdot\bm{e}\,,
\]
where $\bm \theta_{pk}= \bm p_\perp/\ve-\bm k_\perp/\omega$.
Here, we assume that the soft photon direction almost coincides with the charge particle direction. This assumption is obvious because the character angle between an emitted photon and a charged particle is $\theta_{char}\sim m/\ve$.
For such a case, we have
\begin{align}\label{qe:ex:he}
F_{p}^{(+)}(T)=i\int_{T}^{\infty}\frac{dT'}{\varepsilon}e^{i\frac{\omega}{2\varepsilon^{2}}\int_{T}^{T'}dT''(m^{2}+(\varepsilon\bm{\theta}_{pk}-e\bm{A}(T''))^{2})}(\varepsilon\bm{\theta}_{pk}-e\bm{A}(T'))\cdot\bm{e}\,,\nonumber\\
F_{p}^{(-)}(T)=i\int_{-\infty}^{T}\frac{dT'}{\varepsilon}e^{i\frac{\omega}{2\varepsilon^{2}}\int_{T}^{T'}dT''(m^{2}+(\varepsilon\bm{\theta}_{pk}-e\bm{A}(T''))^{2})}(\varepsilon\bm{\theta}_{pk}-e\bm{A}(T'))\cdot\bm{e}\,,
\end{align}
where we use the notations that are often used under these conditions. They differ from ours by replacing  $\phi\leftrightarrow T$.

\item In the case of a plane wave field with frequency $\omega_{0}$ and conditions $p_{-}\omega_{0}\gg pk$, the result is simplified. In this case we can use the average in the period value. Thus, the $\pi_{p}^{\mu}=p^{\mu}+\frac{m^{2}\xi^{2}}{2p_{-}}n^{\mu}$ does not depend on $\phi$, and we can use Eq.\eqref{eq:F_lcf} with $\pi_{p}$ discussed above. Where $\xi=|e| E_0/m\omega_0$, $E_0$, and $\omega_0$ are the laser electric field amplitude and its angular frequency.

\item If the direction of the emitted photon coincides with the laser propagation direction of a laser field $\bm n$, the quantity $k \pi_p(\phi)=p_- \omega$ does not depend on $\phi$ and $\pi_p(\phi)e^*=(p-eA(\phi))e^*$, we have
\begin{align}\label{eq:Fpmcol}
F_{p}^{(-)}(\phi)&=-i\int_{-\infty}^{\phi}\frac{d\varphi}{p_{-}}e^{i(\varphi-\phi)\omega}(p-eA(\phi))e^*\,,\nonumber\\
F_{p}^{(+)}(\phi)&=-i\int_{\phi}^{+\infty}\frac{d\varphi}{p_{-}}e^{i(\varphi-\phi)\omega}(p-eA(\phi))e^*\,,
\end{align}
This result is consistent with \cite{PhysRevD.103.016004}, where the emission of photons collinear with the laser field direction is discussed.
\end{itemize}

\section{Examples}
Here we present several examples of soft photon approximation. The goal of this section is to show how to use the obtained approximation and compare the approximate result with the previously known exact results.

\subsection{Nonlinear Compton scattering}
The nonlinear Compton scattering has been extensively studied, see \cite{Harvey2009,Narozhnyi1996,Ritus1985,PhysRev.133.A705,Dinu2018,Seipt2011,Ivanov2003,Ivanov2003a,Zeldovich,NikiRitus,Baier1975,Mackenroth2011} . The matrix element has the form
\begin{align}
S&=\int d^4x\bar{U}^{(out)}_{p'}(x)e^{ikX}\hat{e}^{*}U^{(in)}_{p}(x)\,,\nonumber
\end{align}
where $p$, $p'$, and $k$ are the momentum of the initial electron, final electron, and emitted photon, respectively. $e^\mu$ is a photon polarization vector.

At first glance, it seems that the soft photon approximation is not applicable for this process. There is no "hard part" of the process. Nevertheless, we can apply this approximation in the following way
\begin{align}
S=&\frac{1}{2}\int d^4x\bar{U}^{(out)}_{p'}(x)[\hat{\Pi}(\Phi)-m] \frac{1}{[\hat{\Pi}(\Phi)]^{2}-m^{2}+i0}[\hat{\Pi}(\Phi)+m]  e^{ikX}\hat{e}^{*}U^{(in)}_{p}(x) \nonumber\\ +
&\frac{1}{2}\int d^4x\bar{U}^{(out)}_{p'}(x) e^{ikX}\hat{e}^{*} [\hat{\Pi}(\Phi)+m] \frac{1}{[\hat{\Pi}(\Phi)]^{2}-m^{2}+i0}[\hat{\Pi}(\Phi)-m] U^{(in)}_{p}(x)=\nonumber\\
=&\frac{1}{2}\int d^4x\bar{U}^{(out)}_{p'}(x)\left([\overset{\rightarrow}{\hat{\Pi}}(\Phi)-m] e^{ikX} F_p^{(-)}(\phi)+F_p^{(+)}(\phi)e^{ikX}[\overset{\leftarrow}{\hat{\Pi}}(\Phi)-m] \right)U^{(in)}_{p}(x)\nonumber \\
=&(2\pi)^3 \delta(\bm p_\perp-\bm p'_\perp-\bm k_\perp) \delta(p_-- p'_-- k_-)\int_{-\infty}^{+\infty}d\phi (\pi_p(\phi) e^*) e^{i\int_{-\infty}^{\phi}\frac{d\varphi'}{p'_{-}}(\pi_{p}(\varphi')k)}\,.
\end{align}
Here we use the symmetric form of the matrix element in order to retain a gauge invariant. We also use the fact that  $F_p^{(-)}$ and $F_{p'}^{(+)}$ have exactly the same phase in the integrand.

The electron-emission  energy spectrum  $\dfrac{d E}{d^3k}$ has the following form
\begin{align}\label{eq:res_CS}
\dfrac{d E}{d^3 k}&=\frac{\alpha m^2}{4\pi^2\omega_0^2 p_-^2}\left[\xi^2(|f_1|^2-\operatorname{Re} \, f_0 f_2^*)-|f_0|^2\right]\,,\\
f_i&=\int_{-\infty}^{+\infty} d\phi\left(\frac{e A(\phi)}{m\xi}\right)^i e^{-i\int_{0}^{\phi}\frac{d\phi'}{p'_{-}}\pi_{p}(\phi')k}\,,\nonumber
\end{align}
where $\xi=|e| E_0/m\omega_0$, $E_0$, and $\omega_0$ are the laser electric field amplitude and its angular frequency.
The obtained energy spectrum for nonlinear Compton scattering \eqref{eq:res_CS} coincides with the classical result \cite{Ritus1985}  if we replace $p'_-\rightarrow p_-$ in the phase of $f_i$. On the other hand, the result \eqref{eq:res_CS} coincides with the exact energy spectrum for nonlinear Compton scattering \cite{Mackenroth2011}, with the leading order of the parameter $\omega/\varepsilon$ in the pre-exponent and exactly in the phase factor.

\begin{figure}
	\centering
	\includegraphics[width=0.9\linewidth]{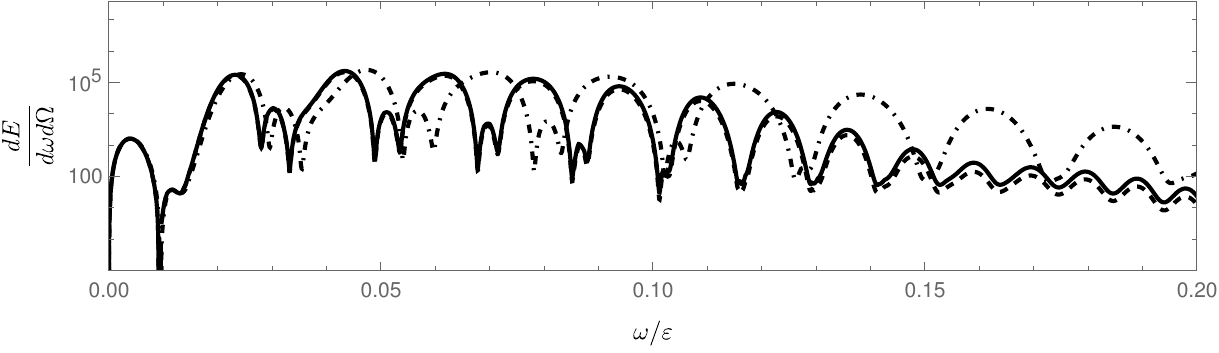}
	\caption{The energy spectrum for nonlinear Compton scattering $\dfrac{d E}{d\omega d\Omega}$ as a function of  $\omega/\varepsilon$, the photon angles are $\theta_x=\frac{m}{\ve}$, $\theta_y=\frac{m}{2\ve}$, the detail of numerical computation discussed in text.   Black line correspond to the exact result, dashed line correspond to the soft photon approximation result, dot-dashed line correspond to the classical result.}
	\label{pic:Compton}
\end{figure}

In order to show the importance of an exact in phase result, we plot the energy spectrum in Fig. \ref{pic:Compton}. For numerical evaluations, we consider electrons with initial electron energy $\varepsilon=10^4 m$ in head-on collisions with the linear polarized laser pulse. Calculations have been performed for a $\xi=1$ and a pulse shape  $\bm A(\phi)= \cos[\omega_0\phi] g(\phi)\bm e_x$, with the envelope function $g(\phi)=\cos^2(\pi \omega_0 \phi/2\tau)$ for $-\tau\geq \phi\geq \tau$ and a zero otherwise, such that is the dimensionless FWHM pulse length  with $\tau=20$ corresponding to 9 fs FWHM for $\omega_0=1.55$ eV.

We see from Fig.~\ref{pic:Compton} that the classical result coincides with the exact one only in the narrow region $\omega/\varepsilon\lesssim0.01$, while the soft photon well fits the exact result in a wider range. The reason for the significant discrepancy between classical and exact quantum results can be simply explained as follows. Although the integrand in the phase differs slightly in the classical and quantum results, the integral in a wide range for a long laser pulse leads to a difference in the phases of the order of unity. The difference in the phases by the order of unity leads to a significant difference in the spectrum. The phase in the soft photon approximation is exactly the same as in the exact result, resulting in significantly better accuracy compared to the classical result.

\subsection{High-energy photon emission in the superposition of a laser and atomic fields}
This process was considered in the different regimes in \cite{Krachkov2019a,PhysRevA.75.053412,PhysRevLett.98.043002,PhysRevA.100.052502}. We consider the special case where high-energy electrons are counter-propagated by a laser field. In such a case, there is an electron wave function in the superposition of a laser and atomic fields \cite{PhysRevA.89.062114}. The "hard process" is the elastic scattering of electrons in the presence of a laser and an atomic field. The amplitude of the elastic scattering can be obtained by the asymptotic  wave function at a large distance \cite{LL4}. The cross section of an elastic scattering $d\sigma_0$ has the form
\begin{align}\label{eq:el_scat}
d\sigma_0&=R(\bm\Delta_\perp) \frac{d\bm p'_\perp}{(2\pi)^2}\,,\nonumber\\
R(\bm\Delta_\perp)&=\int d\bm \rho_1 d\bm \rho_2 e^{-i\bm \Delta_\perp\cdot(\bm\rho_1-\bm\rho_2)}\left[e^{i\mathcal{V}(\rho_{1})-i\mathcal{V}(\rho_2)}-1\right]\,,\nonumber\\
\mathcal{V}(\rho)&=\int_{-\infty}^{\infty} V(\rho,z) dz\,,
\end{align}
where $\Delta=\bm p'-\bm p$ is a momentum transfer, $\bm p$ and $\bm p'$ are the momentum of the initial and final electrons, respectively, and $V(\rho,z)$ is an atomic potential. Note that this cross section is independent of the laser field and coincides with the cross section of the elastic scattering in the atomic field \cite{Krachkov2015}.

Using Eqs. \eqref{qe:ex:he} and \eqref{eq:el_scat} one can straightforwardly obtain the differential cross section of high-energy photon emission in the superposition of a laser and atomic fields
\begin{align}\label{eq:dcs:bs}
d\sigma=&\frac{\alpha}{(2\pi)^4}R(\bm\Delta_\perp)|\bm f_1+\bm g_1|^2\frac{d\bm p'_\perp d\bm k_\perp d\omega}{\omega}\,,\nonumber\\
\bm f_1=&\int_{T}^{\infty}dT'e^{i\frac{\omega}{2\varepsilon(\varepsilon-\omega)}\int_{T}^{T'}dT''(m^{2}+(\varepsilon\bm{\theta}_{p'k}-e\bm{A}(T''))^{2})}(\varepsilon\bm{\theta}_{p'k}-e\bm{A}(T'))\,,\nonumber\\
\bm g_1=&\int_{-\infty}^{T} dT'e^{i\frac{\omega}{2\varepsilon(\varepsilon-\omega)}\int_{T}^{T'}dT''(m^{2}+(\varepsilon\bm{\theta}_{pk}-e\bm{A}(T''))^{2})}(\varepsilon\bm{\theta}_{pk}-e\bm{A}(T'))\,.
\end{align}

The conditions of applicability in such a case are the following: $\omega\ll\varepsilon$ and $\bm k_\perp\ll\bm\Delta_\perp$. Under this assumption, this result is in agreement with \cite{Krachkov2019} (see Eq. (10)).

\section{Factorization of $n$ soft photon emission amplitude}
As mentioned above, the soft photon approximation result coincides with the classical one if we neglect $k_-$ in comparison with $p_-$ in  phase. It means that we neglect retardation. Thus, the emission of several photons should occur independently. 

First, we show that the emission of two photons occurs independently. There are three possibilities. Firstly, one photon is emitted from the initial electron line, while a second photon is emitted from the final electron. In this case, factorization is obvious. The second and third cases correspond to  situations where both photons are emitted from the initial and final electron, respectively.

The factor $F_2^{(-)}(\phi)$ corresponding to the second case (both photons are emitted from the initial electron)  can be found as follows
\begin{align}\label{eq:F2}
&F_2^{(-)}(\phi)U^{(in)}_{p}(x)=\frac{1}{[\hat{\Pi}(\Phi)]-m+i0}\hat{e_2}^{*}e^{ik_2X}\frac{1}{[\hat{\Pi}(\Phi)]-m+i0}\hat{e_1}^{*}e^{ik_1X}U^{(in)}_{p}(x)+(k_1\leftrightarrow k_{2},e_1\leftrightarrow e_{2})=\\\nonumber
&\frac{1}{[\hat{\Pi}(\Phi)]-m+i0}\hat{e_2}^{*}e^{i(k_1+k_2)X}F_{p,k_1}^{(-)}(\phi)U^{(in)}_{p}(x)+(k_1\leftrightarrow k_{2},e_1\leftrightarrow e_{2})\,,
\end{align}
where $ k_i$ and $e_i$ are the momentum and polarization vectors of the emitted photon. To distinguish different photon factors, we use the notation $F_{p,k_1}^{(\pm)}(\phi)$, which corresponds to the $F_{p}^{(\pm)}(\phi)$ from Eq.~\eqref{eq:FGpm}  with the replacement $k\to k_i,e^\mu\to e^\mu_i$. 
Note that the operator $\hat\Pi$ in the last line of Eq.~\eqref{eq:F2} acts not only at $U^{(in)}_{p}(x)$ but also at $F_{p,k_1}^{(-)}(\phi)$.  The result has the following form
\begin{align}\label{eq:F2:1}
&F_2^{(-)}(\phi)=-\int_{-\infty}^{\phi}\frac{d\varphi_{2}}{p_{-}}e^{-i\int_{\varphi_{2}}^{\phi}\frac{d\varphi_{2}'}{p_{-}}(\pi_{p}(\varphi_{2}')(k_{1}+k_{2}))}(\pi_{p}(\varphi_{2}) e_2^*) \int_{-\infty}^{\varphi_{2}}\frac{d\varphi_{1}}{p_{-}}e^{-i\int_{\varphi_{1}}^{\varphi_{2}}\frac{d\varphi_{1}'}{p_{-}}(\pi_{p}(\varphi_{1}')k_{1})}(\pi_{p}(\varphi_{1})e_1^*)+(k_1\leftrightarrow k_{2},e_1\leftrightarrow e_{2})\nonumber\\
&=-\int_{-\infty}^{\phi}\frac{d\varphi_{2}}{p_{-}}e^{-i\int_{\varphi_{2}}^{\phi}\frac{d\varphi_{2}'}{p_{-}}(\pi_{p}(\varphi_{2}')k_{2})}(\pi_{p}(\varphi_{2}) e_2^*) \int_{-\infty}^{\varphi_{2}}\frac{d\varphi_{1}}{p_{-}}e^{-i\int_{\varphi_{1}}^{\phi}\frac{d\varphi_{1}'}{p_{-}}(\pi_{p}(\varphi_{1}')k_{1})}(\pi_{p}(\varphi_{1})e_1^*)+(k_1\leftrightarrow k_{2},e_1\leftrightarrow e_{2})\,,
\end{align}
By changing the order of integration in the second term, we obtain that factorization takes place
\begin{align}\label{eq::F2_final}
F_2^{(-)}(\phi)=-\int_{-\infty}^{\phi}\frac{d\varphi_{2}}{p_{-}}e^{-i\int_{\varphi_{2}}^{\phi}\frac{d\varphi_{2}'}{p_{-}}(\pi_{p}(\varphi_{2}')k_{2})}(\pi_{p}(\varphi_{2}) e_2^*) \int_{-\infty}^{\phi}\frac{d\varphi_{1}}{p_{-}}e^{-i\int_{\varphi_{1}}^{\phi}\frac{d\varphi_{1}'}{p_{-}}(\pi_{p}(\varphi_{1}')k_{1})}(\pi_{p}(\varphi_{1})e_1^*)\,.
\end{align}
Factor $F_2^{(+)}(\phi)$, which corresponds to the case where both photons are emitted by the final electron,  possesses the analogous factorization.  The final result has the following form
$$
S=\int d\phi \tilde S_0(\phi) \left(F_{p,k_1}^{(-)}(\phi)+F_{p',k_1}^{(+)}(\phi)\right)\left(F_{p,k_2}^{(-)}(\phi)+F_{p',k_2}^{(+)}(\phi)\right)\,,
$$

Note that here we neglected $k_{i,-}$ in comparison with $p_-$ in the phase denominator, which corresponds to  the classical result. Without this simplification, there is no factorization. Thus, the emission of two soft photons occurs independently. This method is trivially generalized by the case of $n$ photon emission, where the same factorization takes place. For the $n$-photon emission factor of the initial electron, we have
\begin{align}
&F_n^{(-)\lambda_{1}..,\lambda_{n}}(\phi)=\sum_{\sigma\in\mathcal{S}_n}\underset{\phi>\varphi_{\sigma(1)}>...\varphi_{\sigma(n)}>-\infty}{\idotsint}\Pi_{i=1}^{i=n}\,d\varphi_{i} f^{(-)}_{i}(\varphi_{i})\nonumber\\
&=\Pi_{i=1}^{i=n}\int_\infty^\phi\,d\varphi_{i} f^{(-)}_i(\varphi_{i})\sum_{\sigma\in\mathcal{S}_n} H(\phi-\varphi_{\sigma(1)})H(\varphi_{\sigma(1)}-\varphi_{\sigma(2)})...H(\varphi_{\sigma(n-1)}-\varphi_{\sigma(n)})\nonumber\\
&=\Pi_{i=1}^{i=n}\left[\int_\infty^\phi\,d\varphi_{i} f^{(-)}_i(\varphi_{i})\right]\,,
\end{align}
where the sum is over all permutations $\sigma$ of $n$ elements, $H(x)$ is a Heaviside step function, and $f^{(-)}_i(\varphi)=-i\frac{(\pi_{p}(\varphi)e^*_i)}{p_{-}}e^{-i\int_{\varphi}^{\phi}\frac{d\varphi'}{p_{-}}(\pi_{p}(\varphi')k_i)}$ is an integrand of $F_{p,k_i}^{(-)}(\phi)$. Thus, factorization takes place, and the emission of $n$ soft photons occurs independently. The matrix element for process $S$ with $n$ additional photons has the form

$$
S=\int d\phi \tilde S_0(\phi) \Pi_{i=1}^{i=n} \left(F_{p,k_i}^{(-)}(\phi)+F_{p',k_i}^{(+)}(\phi)\right)\,.
$$

\section{Cancellations of infrared divergences in the presence of a plane wave laser field}

\begin{figure}[h]
	\begin{subfigure}{\textwidth}
		\includegraphics[width=0.58\linewidth]{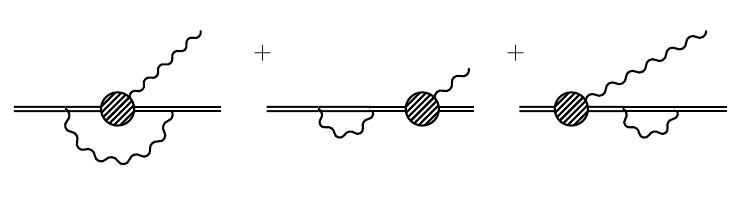}
		\caption{Virtual radiative correction diagrams.}
		\label{fig:virt}
	\end{subfigure}
	\begin{subfigure}{\textwidth}
		\centering\includegraphics[width=0.38\linewidth]{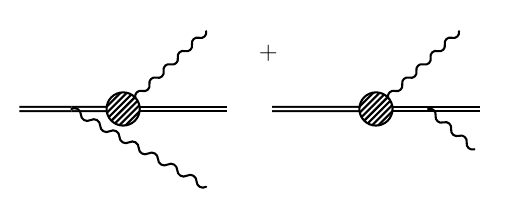}
		\caption{Real radiative correction diagrams.}
		\label{fig:real}
	\end{subfigure}
	\caption{Feynman diagrams corresponding to the Born and  one-loop radiative corrections  to the nonlinear Compton scattering.
	}
	\label{fig:diagrams}
\end{figure}

As is well known \cite{Bloch1937}, the infrared divergence term from the virtual radiative  corrections for arbitrary processes must be cancelled by the processes with additional soft photon in QED. In this chapter, we will show that this statement holds true for QED with the plane wave laser field background. The infrared divergence in the laser field was previously discussed in \cite{dinu2012infrared,ilderton2013scattering,di2018analytical,ilderton2020analytic}.

We consider the soft part of radiative correction for nonlinear Compton scattering. The Feynman diagrams of a vitrual and real radiative correction  are depicted in Fig. \ref{fig:diagrams}. We will show that total probability is infrared finite, i.e. the  interference between Born  and virtual radiative correction matrix element (see Fig.~\ref{fig:virt}) cancels the infrared divergence of real radiative correction (see Fig.~\ref{fig:real}).

We start our computation with the vertex correction. For the plane wave field background, it was discussed in \cite{PhysRevD.102.076018}. We consider the soft part of the radiative corrections and restrict the integral over $\bm k$ in the following way: $\lambda<|\bm k|<\Lambda$.
The infrared cutoff $\Lambda$ is some convenient dividing point chosen low enough to satisfy the approximations 
made above in Section \ref{sec::ampl}. In addition,  we will also impose a cutoff $\lambda$ in order to display the logarithmic divergences as powers of $\log\lambda$. We take $\lambda$ very small in particular, $\lambda\ll\Lambda$ so this cutoff only affects the infrared lines because it is only these that give infrared divergences for $\lambda=0$.

The soft part of the one loop vertex correction matrix element for  nonlinear Compton scattering can be computed by soft photon approximation. The matrix element integrand over $k$ corresponds to the emission of soft photon from the initial electron line with momentum $k$ and absorption of this photon  from the final electron line, which corresponds to the  emission of soft photon from the final electron line with momentum $-k$. It has the following form
\begin{align}\label{eq::rad::virt}
S_{vertex}=&ie^2  \int d\phi (2\pi)^3 \delta(\bm p_\perp-\bm p'_\perp-\bm q_\perp) \delta(p_-- p'_-- q_-)  M_0(\phi)
\int_\lambda^\Lambda\frac{d^4 k}{(2\pi)^4}\frac{1}{k^2+i0}\nonumber\\
\times&\int_{-\infty}^{\phi}\frac{d\varphi_1}{p_{-}}e^{-i\int_{\varphi_1}^{\phi}\frac{d\varphi_1'}{p_{-}-k_-}(\pi_{p}(\varphi_1')k-i 0)}\pi^\nu_{p}(\varphi_1)
\int_{\phi}^{+\infty}\frac{d\varphi_2}{p'_{-}}e^{-i\int_{\phi}^{\varphi_2}\frac{d\varphi_2'}{p'_{-}-k_-}(\pi_{p'}(\varphi_2')k-i 0)}\pi_{p',\nu}(\varphi_2)\,,
\end{align}
where  $M_0(\phi)$ is an integrand over $\phi$ of the matrix element of single nonlinear Compton scattering
\begin{align}\label{eq::M00}
e\int d^4x\bar{U}^{(out)}_{p'}(x)e^{iqX}\hat{e}^{*}U^{(in)}_{p}(x)= \int d\phi (2\pi)^3 \delta(\bm p_\perp-\bm p'_\perp-\bm q_\perp) \delta(p_-- p'_-- q_-)  M_0(\phi)\,.
\end{align}

The limits on the integral in \eqref{eq::rad::virt} refer to $|\bm k|$. The integrand of  \eqref{eq::rad::virt} is analytic in $k^0$ except at the four poles. Only one pole $k^0=|\bm k|-i0$  is in the lower half-plane. We close the $k^0$ contour with a large semicircle in  the lower half-planes and compute this integral by residue
\begin{align}\label{eq::rad::virt1}
S_{vertex}=&e^2  \int d\phi (2\pi)^3 \delta(\bm p_\perp-\bm p'_\perp-\bm q_\perp) \delta(p_-- p'_-- q_-)  M_0(\phi)
\int_\lambda^\Lambda\frac{d^3 k}{(2\pi)^3 2|\bm k|}\nonumber\\
\times&\int_{-\infty}^{\phi}\frac{d\varphi_1}{p_{-}}e^{-i\int_{\varphi_1}^{\phi}\frac{d\varphi_1'}{p_{-}-k_-}(\pi_{p}(\varphi_1')k)}\pi^\nu_{p}(\varphi_1)
\int_{\phi}^{+\infty}\frac{d\varphi_2}{p'_{-}}e^{-i\int_{\phi}^{\varphi_2}\frac{d\varphi_2'}{p'_{-}-k_-}(\pi_{p'}(\varphi_2')k)}\pi_{p',\nu}(\varphi_2)\,.
\end{align}

Below, we will omit $k_-$ compared to $p_-\,,p'_-$ in phase. Without this simplification, there is no factorization of matrix element and no cancellation between real and virtual radiative corrections.

Another matrix element depicted in Fig.~\ref{fig:virt} can be obtained in a similar way. For example, the matrix element, which corresponds to the second diagram in Fig.~\ref{fig:virt} can be obtained by $F_2^{(-)}(\phi)$, see Eq.~\eqref{eq::F2_final} by replacing $k_1\to k\,,k_2\to -k$ with the additional factor $\frac{1}{2}$. The factor $\frac{1}{2}$  is required because in  $F_2^{(-)}(\phi)$ computation, we took into account the sum of two diagrams, whereas in this case there is only one diagram.

Matrix element of virtual radiative correction  $S_{virt}$ has the following form
\begin{align}\label{eq::rad::virt1}
S_{virt}=&\frac{e^2}{2}  \int d\phi (2\pi)^3 \delta(\bm p_\perp-\bm p'_\perp-\bm q_\perp) \delta(p_-- p'_-- q_-)  M_0(\phi)
\int_\lambda^\Lambda\frac{d^3 k}{(2\pi)^3 2|\bm k|}\nonumber\\
\times&\left|\int_{-\infty}^{\phi}\frac{d\varphi_1}{p_{-}}e^{-i\int_{\varphi_1}^{\phi}\frac{d\varphi_1'}{p_{-}}(\pi_{p}(\varphi_1')k)}\pi^\nu_{p}(\varphi_1)+
\int_{\phi}^{+\infty}\frac{d\varphi_2}{p'_{-}}e^{i\int_{\phi}^{\varphi_2}\frac{d\varphi_2'}{p'_{-}}(\pi_{p'}(\varphi_2')k)}\pi_{p'}^\nu(\varphi_2)\right|^2\,,
\end{align} 
where we neglect $k_-$ in  comparison with $p_-$ in the phase denominator. The cancellation of infrared divergences takes a place only under this assumption.

The contribution to the probability $dW_{virt}$ from virtual radiative correction has the following form
\begin{align}\label{eq::rad::virt2}
dW_{virt}=&e^2\operatorname{Re}  \int_\lambda^\Lambda\frac{d^3 k}{(2\pi)^3 2|\bm k|}
\int d\phi d\phi'  M_0(\phi) M_0^*(\phi') d\rho_f \nonumber\\
\times&\left|\int_{-\infty}^{\phi}\frac{d\varphi_1}{p_{-}}e^{-i\int_{\varphi_1}^{\phi}\frac{d\varphi_1'}{p_{-}}(\pi_{p}(\varphi_1')k)}\pi^\nu_{p}(\varphi_1)+
\int_{\phi}^{+\infty}\frac{d\varphi_2}{p'_{-}}e^{i\int_{\phi}^{\varphi_2}\frac{d\varphi_2'}{p'_{-}}(\pi_{p'}(\varphi_2')k)}\pi_{p'}^\nu(\varphi_2)\right|^2\,,
\end{align}
where  $d\rho_f$ is  element of the final particle phase space for nonlinear Compton scattering, including $\delta$-functions.

The real radiative correction, which corresponds to the double Compton scattering, has the following form 
\begin{align}\label{eq::rad::real}
dW_{real}=&-e^2  \int_\lambda^E\frac{d^3 k}{(2\pi)^3 2|\bm k|}
\int d\phi d\phi'  M_0(\phi) M_0^*(\phi') d\rho_f \nonumber\\
\times&\left(\int_{-\infty}^{\phi}\frac{d\varphi_1}{p_{-}}e^{-i\int_{\varphi_1}^{\phi}\frac{d\varphi_1'}{p_{-}}(\pi_{p}(\varphi_1')k)}\pi^\nu_{p}(\varphi_1)+
\int_{\phi}^{+\infty}\frac{d\varphi_2}{p'_{-}}e^{i\int_{\phi}^{\varphi_2}\frac{d\varphi_2'}{p'_{-}}(\pi_{p'}(\varphi_2')k)}\pi_{p'}^\nu(\varphi_2)\right) \nonumber\\
\times&\left(\int_{-\infty}^{\phi'}\frac{d\varphi_1}{p_{-}}e^{-i\int_{\varphi_1}^{\phi'}\frac{d\varphi_1'}{p_{-}}(\pi_{p}(\varphi_1')k)}\pi^\nu_{p}(\varphi_1)+
\int_{\phi'}^{+\infty}\frac{d\varphi_2}{p'_{-}}e^{i\int_{\phi'}^{\varphi_2}\frac{d\varphi_2'}{p'_{-}}(\pi_{p'}(\varphi_2')k)}\pi_{p'}^\nu(\varphi_2)\right)^*\,,
\end{align}
here we assume that the energy of the soft photon less than the detector's efficiency $E$ and greater than the infrared cutoff $\lambda$.

Note that for values of small $k$ the integral over $\varphi_{1,2}$ converges at large distance $|\phi-\varphi_{1,2}|\sim \varphi_{char}$, where $\varphi_{char}$ is the formation length of soft photon emission, while $\phi-\phi'\sim l_f$, where $l_f$ is the formation length of nonlinear Compton scattering \cite{BAIER2005261,Ritus1985}. For small $\omega$ the formation length of a nonlinear Compton scattering is much smaller than the formation length of soft photon emission.  For this reason, we can change $\phi'\rightarrow\phi$ in the last line of Eq.~\eqref{eq::rad::real}. Then the sum of the real and virtual corrections becomes finite and has the following form
\begin{align}\label{eq::rad::summ}
dW_{virt}+ dW_{real}=&e^2 \int_E^\Lambda\frac{d^3 k}{(2\pi)^3 2|\bm k|}
\int d\phi d\phi'  M_0(\phi) M_0^*(\phi') d\rho_f \nonumber\\
\times&\left|\int_{-\infty}^{\phi}\frac{d\varphi_1}{p_{-}}e^{-i\int_{\varphi_1}^{\phi}\frac{d\varphi_1'}{p_{-}}(\pi_{p}(\varphi_1')k)}\pi^\nu_{p}(\varphi_1)+
\int_{\phi}^{+\infty}\frac{d\varphi_2}{p'_{-}}e^{i\int_{\phi}^{\varphi_2}\frac{d\varphi_2'}{p'_{-}}(\pi_{p'}(\varphi_2')k)}\pi_{p'}^\nu(\varphi_2)\right|^2\nonumber\\
\end{align}

Note that the last line in Eq.~\eqref{eq::rad::summ} is a $-\sum_\lambda \left|F_{p}^{(-)}(\phi)+F_{p'}^{(+)}(\phi)\right|^2$, where the sum is taken by the photon polarization. Thus, this answer can be trivially generalized by arbitrary processes
\begin{align}\label{eq::rad::gen}
dW_{rad}=\int d\phi \frac{dW_{B}}{d\phi} (-e^2) \int_E^\Lambda\frac{d^3 k}{(2\pi)^3 2|\bm k|} \sum_\lambda\left|\sum_{i} F_{p_i}^{(-)}(\phi)+\sum_{f} F_{p_f}^{(+)}(\phi)\right|^2\,,
\end{align}
where $dW_{B}$ is  probabilities in the Born approximation and $i\,,f$ is a set of charged initial and final particles. Note that in the case of antiparticles, we should use $G$ instead of $F$.

\section{Conclusion}\label{conclusion}
In the present paper, we consider the processes involving the emission of soft photons in the presence of a strong laser field. We show that the matrix element $S$ of the process  $\text{i}\rightarrow\text{f}+\gamma$ with a soft photon $\gamma$ can be expressed in terms of the matrix element $S_0$ of the process $\text{i}\rightarrow\text{f}$. The recipe for using the soft photon approximation is as follows. To obtain the amplitude $S$ from the amplitude $S_0$, we should multiply the integrand $S_0$ with respect to the variable $\phi$ by the sum of the factors \eqref{eq:FGpm}, one for each outgoing or incoming charged particle. These factors are universal for a generic plane wave laser field background and do not depend on the spin of the particle. This approximation allows us to obtain a result that is exact in the phase and approximate in the prefactor to order $\mathcal{O}(\omega/\varepsilon_\text{char})$, where $\omega$ is the frequency of the soft photon and $\varepsilon_\text{char}$ is the characteristic energy of the $\text{i}\rightarrow\text{f}$ process.

We present several important applications of this soft photon approximation. We compute the probabilities of nonlinear Compton scattering and photon emission in the superposition of a laser and atomic fields by soft photon approximation and compare obtained result with the exact one.  We demonstrate that the amplitude of $n$ soft photons emission has factorization, which corresponds to the independence of the emission of $n$ soft photons. Third, we use the discussed approximation to prove cancellation of real and virtual infrared divergences for nonlinear Compton scattering and derive the finite radiative corrections.

The soft photon approximation is a useful tool for QED studies in the presence of a strong laser field. For example, it can be used for nonlinear double Compton scattering in the case of one or both photon are soft. It may helps also to obtain the result for $e^+e^-\to\gamma\gamma$, in the case of one photon is soft. Besides, soft photon approximation may be widely used for computation of infrared part of radiative correction.

\section*{Appendix}\label{sec:app} 
This appendix provides additional details on the formulas and notation used in the paper.

Let vector $\bm{n}$ define the propagation direction of the plane wave. Vector potential $A^{\mu}(\phi)$ depends only on  $\phi=t-\bm{n}\cdot\bm{x}.$ We introduce four four-dimensional quantities: $n^{\mu}=(1,\bm{n})$,
$\tilde{n}^{\mu}=(1,-\bm{n})/2$, and $a_{j}^{\mu}=(0,\bm{a}_{j})$, where $j=1,2$. The four-dimensional quantities $n^{\mu}$, $\tilde{n}^{\mu}$, and $a_{j}^{\mu}$
fulfill the completeness relation  
\begin{eqnarray*}
	\eta^{\mu\nu} & = & n^{\mu}\tilde{n}^{\nu}+\tilde{n}^{\mu}n^{\nu}-a_{1}^{\mu}a_{1}^{\nu}-a_{2}^{\mu}a_{2}^{\nu}\,.
\end{eqnarray*}
Note that $(n\tilde{n})=1$, $(n a_{j})=(\tilde{n}a_{j})=\tilde{n}^{2}=n^{2}=0$,
$a_{i}a_{j}=-\delta_{i,j}$. In what follows, we refer to the longitudinal ($n$) direction as the direction along $\bm{n}$ and to the transverse ($\perp$) plane as the plane spanned by the two perpendicular unit vectors $\bm{a}_{j}$. Coordinates:
\[
\phi=(nx)=t-x_{n}\,,\quad T=(\tilde{n}x)=(t+x_{n})/2\,,\quad\bm{x}_{\perp}^{i}=-(a_{i}x)\,,
\]

where $x_{n}=\bm{n}\cdot\bm{x}$. For an arbitrary four-vector $p$, we
introduce $p_{-}=(np)=p^{0}-p_{n}$, $p_{+}=(\tilde{n}p)=(p^{0}+p_{n})/2\,$,
and $\bm{p}_{\perp}=(p_{\perp,1},p_{\perp,2})=-((pa_{1}),(pa_{2}))=(\bm{p}\cdot\bm{a}_{1},\bm{p}\cdot\bm{a}_{2})$.
Thus, 
\[
px=p_{\mu}x_{\nu}\eta^{\mu\nu}=(xn)(p\tilde{n})+(pn)(x\tilde{n})-\bm{x}_{\perp}\cdot\bm{p}_{\perp}=p_{-}T+p_{+}\phi-\bm{x}_{\perp}\cdot\bm{p}_{\perp}\,.
\]

The momenta operators on this basis have the form
\[
\begin{array}{c}
P_{\phi}=-i\partial_{\phi}=-(\tilde{n}P)=-(i\partial_{t}-i\partial_{x_{n}})/2\quad P_{T}=-i\partial_{T}=-(nP)=-(i\partial_{t}+i\partial_{x_{n}}),\\
\bm{P}_{\perp}=(P_{\perp,1},P_{\perp,2})=-i(\bm{a}_{1}\cdot\bm{\nabla},\bm{a}_{2}\cdot\bm{\nabla})\,.
\end{array}
\]
They satisfy the following commutation relation:
\[
[\phi,P_{\phi}]=[T,P_{T}]=i\,,\quad[X_{\perp,j},P_{\perp,k}]=i\delta_{jk}\,,
\]

which are equivalent to the commutation relations $[X^{\mu},P^{\nu}]=-i\eta^{\mu\nu}$,
with $P^{\mu}=i\partial^{\mu}$.

We will need the following identities  
\begin{align}\label{eq:id1}
\exp(i(Xq))g(P)\exp(-i(Xq)) & =g(P+q),\nonumber\\
\exp(i(Py))f(X)\exp(-i(Py)) & =f(X-y),
\end{align}
where $q^{\mu}$ and $y^{\mu}$ are constant four-vectors.

In addition, the commutation relations $[\phi,P_{\phi}]=[T,P_{T}]=i$ imply, in particular, the identities 
\begin{align}\label{eq:id2}
\exp(ia\phi)\tilde{g}(P_{\phi})\exp(-ia\phi) & =\tilde{g}(P_{\phi}-a),\nonumber\\
\exp(ibP_{\phi})\tilde{f}(\phi)\exp(-ibP_{\phi}) & =\tilde{f}(\phi+b),
\end{align}
with $a$ and $b$ being two constants and $\tilde{f}(\phi)$ and
$\tilde{g}(P_{\phi})$ being two arbitrary functions. For $T,P_{T}$, the identities are the same.

The Volkov states $U_{p}(x)$ and $V_{p}(x)$ can be classified
by means of the asymptotic momentum quantum numbers $\bm{p}$ (and
then the energy $\varepsilon=\sqrt{m^{2}+\bm{p}^{2}}$) and of the
asymptotic spin quantum number $s$ in the remote past, i.e., for
$t\rightarrow-\infty$ for $(in)$-state and for remote future for $(out)$-state.  Following the general notation in Ref.~\cite{LL4}, these states can be written as
\begin{align}\label{eq:vawe_func}
U_{p}^{(in)}(x)&=\bigg[1+\frac{e\hat{n}\hat{A}(\phi)}{2p_{-}}\bigg]u_{p}\text{e}^{i\left\{ -(px)-\int_{-\infty}^{\phi}d\varphi\left[\frac{e(pA(\varphi))}{p_{-}}-\frac{e^{2}A^{2}(\varphi)}{2p_{-}}\right]\right\} }\,,\nonumber\\
\bar{U}_{p}^{(out)}(p,x)&=\bar{u}{}_{p}\bigg[1-\frac{e\hat{n}\hat{A}(\phi)}{2p_{-}}\bigg]\text{e}^{i\left\{ (px)-\int_{\phi}^{+\infty}d\varphi\left[\frac{e(pA(\varphi))}{p_{-}}-\frac{e^{2}A^{2}(\varphi)}{2p_{-}}\right]\right\} }\,,\nonumber\\
V_{p}^{(out)}(x)&=\bigg[1-\frac{e\hat{n}\hat{A}(\phi)}{2p_{-}}\bigg]v_{p}\text{e}^{i\left\{ (px)+\int_{\phi}^{+\infty}d\varphi\left[\frac{e(pA(\varphi))}{p_{-}}+\frac{e^{2}A^{2}(\varphi)}{2p_{-}}\right]\right\} }\,,\nonumber\\
\bar{V}_{p}^{(in)}(x)&=\bar{v}{}_{p}\bigg[1+\frac{e\hat{n}\hat{A}(\phi)}{2p_{-}}\bigg]\text{e}^{i\left\{ -(px)+\int_{-\infty}^{\phi}d\varphi\left[\frac{e(pA(\varphi))}{p_{-}}+\frac{e^{2}A^{2}(\varphi)}{2p_{-}}\right]\right\} }\,,
\end{align}
where $u_{p}$ and $v_{p}$ are the free spinors. Here we have introduced the notation $\hat{v}=\gamma^{\mu}v_{\mu}$
for a generic four-vector $v^{\mu}$, with $\gamma^{\mu}$ being the Dirac matrices.

The electron Green's function $G(x,x')$ in the general plane-wave
background electromagnetic field described by the four-vector potential
$A^{\mu}(\phi)$ is defined by the equation 
\begin{equation}
\{\gamma^{\mu}[i\partial_{\mu}-eA_{\mu}(\phi)]-m\}G(x,x')=\delta^{(4)}(x-x').
\end{equation}
Here, we always assume the Feynman prescription corresponding to the shift $m\rightarrow m-i0$ \cite{LL4}. Within
the operator technique, the operator $G$ corresponding to the Green's
function $G(x,x')$ is defined via the equation $G(x,x')=\langle x|G|x'\rangle$,
i.e., as 
\begin{equation}
G=\frac{1}{\hat{\Pi}-m+i0}=(\hat{\Pi}+m)\frac{1}{\hat{\Pi}^2-m^2+i0}=\frac{1}{\hat{\Pi}^2-m^2+i0}(\hat{\Pi}+m)\,,
\end{equation}
where $\Pi^{\mu}=P^{\mu}-eA^{\mu}(\Phi)$. In \cite{Baier1975} was shown that squared Green's function can be written in the form 
\begin{align}
&\frac{1}{\hat{\Pi}^{2}-m^{2}+i0}=(-i)\int_{0}^{\infty}ds\,e^{-im^{2}s}e^{2isP_{T}P_{\phi}} e^{-i\int_{0}^{s}ds'[\bm{P}_{\perp}-e\bm{A}_{\perp}(\Phi-2s'P_{T})]^{2}}\Big\{1-\frac{e}{2P_{T}}\hat{n}[\hat{A}(\Phi-2sP_{T})-\hat{A}(\Phi)]\Big\},\nonumber\\
&=(-i)\int_{0}^{\infty}du\,e^{-im^{2}u}\Big\{1+\frac{e}{2P_{T}}\hat{n}[\hat{A}(\Phi+2uP_{T})-\hat{A}(\Phi)]\Big\} e^{-i\int_{0}^{u}du'[\bm{P}_{\perp}-e\bm{A}_{\perp}(\Phi+2u'P_{T})]^{2}}e^{2iuP_{T}P_{\phi}}.\label{eq:G}
\end{align}
where the prescription $m^{2}\rightarrow m^{2}-i0$ is understood.

\bibliography{sphoton}
\end{document}

%% file: commands.tex
\newcommand{\ve}{\varepsilon}